\documentclass{article}

\usepackage{microtype}
\usepackage{graphicx}
\usepackage{subfigure}
\usepackage{caption}
\usepackage{booktabs} 

\usepackage[unicode=true,pdfusetitle,%
 pdfauthor={Ioanna Bouri, Fanni Franssila, Markku Alho, Giulia Cozzani, Ivan Zaitsev, Minna Palmroth, Teemu Roos},%
 bookmarks=true,bookmarksnumbered=true,bookmarksopen=false,%
 breaklinks=true,pdfborder={0 0 0},backref=false,colorlinks=true,citecolor=blue]{hyperref}

\usepackage[accepted]{icml2023}

\usepackage{amsmath}
\usepackage{amssymb}
\usepackage{mathtools}
\usepackage{amsthm}

\usepackage[inline]{enumitem}

\usepackage[capitalize,noabbrev]{cleveref}

\theoremstyle{plain}

\theoremstyle{definition}

\theoremstyle{remark}

\usepackage[textsize=tiny]{todonotes}

\icmltitlerunning{Graph Representation of the Magnetic Field Topology in High-Fidelity Plasma Simulations for ML Applications}

\begin{document}

\twocolumn[
\icmltitle{Graph Representation of the Magnetic Field Topology in High-Fidelity Plasma Simulations for Machine Learning Applications}



\icmlsetsymbol{equal}{*}

\begin{icmlauthorlist}
\icmlauthor{Ioanna Bouri}{cs}
\icmlauthor{Fanni Franssila}{cs}
\icmlauthor{Markku Alho}{phy}
\icmlauthor{Giulia Cozzani}{phy}
\icmlauthor{Ivan Zaitsev}{phy}
\icmlauthor{Minna Palmroth}{phy}
\icmlauthor{Teemu Roos}{cs}
\end{icmlauthorlist}

\icmlaffiliation{phy}{Department of Physics, University of Helsinki, Helsinki, Finland}
\icmlaffiliation{cs}{Department of Computer Science, University of Helsinki, Helsinki, Finland}

\icmlcorrespondingauthor{Ioanna Bouri}{\href{mailto:ioanna.bouri@helsinki.fi}{ioanna.bouri@helsinki.fi}}

\icmlkeywords{Machine Learning, ICML}

\vskip 0.3in
]



\printAffiliationsAndNotice{}  

\begin{abstract}

Topological analysis of the magnetic field in simulated plasmas allows the study of various physical phenomena in a wide range of settings. One such application is magnetic reconnection, a phenomenon related to the dynamics of the magnetic field topology, which is difficult to detect and characterize in three dimensions. We propose a scalable pipeline for topological data analysis and spatiotemporal graph representation of three-dimensional magnetic vector fields. We demonstrate our methods on simulations of the Earth's magnetosphere produced by Vlasiator, a supercomputer-scale Vlasov theory-based simulation for near-Earth space. The purpose of this work is to challenge the machine learning community to explore graph-based machine learning approaches to address a largely open scientific problem with wide-ranging potential impact.

\end{abstract}

\section{Introduction} 

Magnetic reconnection is a fundamental plasma physical process characterized by a topological reconfiguration of the magnetic field and energy conversion from magnetic to kinetic and thermal energy, leading to plasma heating, particle acceleration, and mixing of plasmas \cite{priest_magnetic_2000, li2021}. The phenomenon is encountered in different settings and plays a key role in the eruption of solar flares and coronal mass ejections (CMEs) in the solar corona \cite{mei_numerical_2012}, in the Earth's magnetosphere and its interaction with the solar wind \cite{angelopoulos2008,palmroth2017,li2022}, in astrophysical plasmas \cite{parker1979,yamada2010}, as well as in fusion plasma during major and minor tokamak disruptions \cite{taylor1986, yamada1999}.

Magnetic reconnection is linked to \emph{space weather} conditions that can potentially damage terrestrial technological infrastructure, satellites, and manned space missions \cite{odenwald2006}. 
CMEs cause magnetospheric magnetic storms \cite{schwenn_association_2005}, during which the terrestrial power grids may suffer from Geomagnetically Induced Currents (GICs) and even fail \cite{pulkkinen_geomagnetic_2005}. Solar flares accelerate particles into relativistic energies, which propagate to the Earth's upper atmosphere and 
affect satellite and radar signals that can be significantly altered or lost during active space weather conditions \cite{baker2004}. 



The nature of the phenomenon is well-understood in two-dimensional (2D) settings, and quasi-2D models have been successful at reproducing many features of reconnection in the solar corona and the Earth's magnetosphere \cite{cassak_scaling_2007, liu_first_principles_2022}.
However, magnetic reconnection is intrinsically a three-dimensional (3D) process. This becomes especially evident when considering reconnection in the solar corona, where the magnetic field forms twisted coronal loops with complex topologies \cite{lee_dimensionality_2022}. Despite considerable progress, the additional complexity introduced in 3D settings continues to pose many open questions regarding the nature of 3D magnetic reconnections in the solar and the Earth's magnetospheric environment \cite{daughton_role_2011}.




We present a scalable pipeline for topological data analysis and graph representation of 3D magnetic vector fields. First, we introduce spatial null graphs, a graph representation that can be used to characterize the topology of a magnetic field. In addition, to encode the temporal evolution of the magnetic field, we extend this concept with spatiotemporal null graphs. Finally, we present the spatial and spatiotemporal null graphs produced by the topological analysis of the magnetic vector field in the Earth's magnetotail. For this purpose, we use 3D global simulations produced by Vlasiator, a supercomputer-scale Vlasov theory-based model that incorporates the solar wind -- magnetosphere -- ionosphere system
\cite{palmroth_vlasov_2018}. The constructed graphs enable the use of topological information as input for machine learning methods such as (spatiotemporal) graph neural networks (GNNs) \cite{reinhart2018, ma_tang_2021}.

\section{Magnetic Field Topology}

This section introduces some concepts of vector field topology. For a general introduction, see~\cite{Gun20b}; for reviews focused on magnetic fields and magnetic reconnection in particular, see~\cite{longcope2005,parnell2010}.  

The magnetic field on a location with spatial coordinates $\vec x =~(x,y,z) \in~\mathbb{R}^3$ can be represented as a vector field $\vec B(\vec x) =~(B_x(\vec x), B_y(\vec x), B_z(\vec x))$. According to Gauss's law for magnetism, the field has zero divergence everywhere
\begin{equation}
\nabla \cdot \vec B
    = \left(\frac{\partial B_x}{\partial x} +
    \frac{\partial B_y}{\partial y} +
    \frac{\partial B_z}{\partial z}\right)
    \equiv 0.
    \label{eq:solenoid}
\end{equation}
From a topological perspective, \emph{magnetic nulls}, points where the magnetic field vanishes $\|\vec B(\vec x_0)\|_2 = 0$, are of special interest. At such points, the structure of the local field can be characterized by forming a first-order Taylor approximation around $\vec x_0$:
$$\vec B(\vec x) =
    J(\vec x_0) (\vec x - \vec x_0)
    + o(\|\vec x - \vec x_0\|_2),
$$
where $J = \nabla \vec B$ is the Jacobian of the magnetic field. 
The topology of the field is characterized by the magnetic skeleton, which comprises of the magnetic nulls, separatrix surfaces delineating distinct magnetic domains, and separator curves formed on the intersections of separatrix surfaces \cite{bungley:1996,longcope2002,longcope2005, parnell2010}. 

To extract the magnetic skeleton of a field, we use the Visualization Toolkit (VTK) \citep{schroeder1998} -- an open source software package for scientific data analysis and visualization. 
The VTK vector field topology filter \citep{bujack2021} is a later extension to the package that adds the functionality for computing the main elements of the topological skeletons of 2D and 3D vector fields.



\subsection{Magnetic nulls}
\label{sec:sec31}
Magnetic nulls can be classified into different types that characterize their topology, according to the eigenvalues of the Jacobian matrix $J$ of the vector field~\cite{lau1990, Gun20b}. 

Given the three eigenvalues of the Jacobian, $(\lambda_1, \lambda_2,\lambda_3) \in \mathbb{C}^3$, it follows from Eq.~\eqref{eq:solenoid} that their sum is equal to zero:
$\lambda_1 + \lambda_2 + \lambda_3 = 0$,
and, therefore, 
two of the eigenvalues must have the same sign while the third one is of the opposite sign\footnote{We do not consider \emph{degenerate nulls} where one or more of the eigenvalues is exactly zero. Such points are physically unstable~\cite{priest:1996} and can be handled as special cases of one or more of the four types we introduce here.}. Moreover, while the eigenvalues can be complex-valued, the eigenvalues with non-zero imaginary parts always come in pairs of complex conjugates, so that their real parts are the same, and the third eigenvalue is a real number of the opposite sign.

\begin{figure}[t]
\centering
\subfigure[Type A]{
\includegraphics[width=1.47cm]{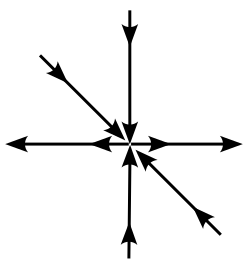}
\label{fig:A}
}
\quad
\subfigure[Type B]{
\includegraphics[width=1.47cm]{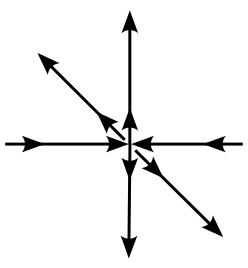}
\label{fig:B}
}
\quad
\subfigure[Type As]{
\includegraphics[width=1.47cm]{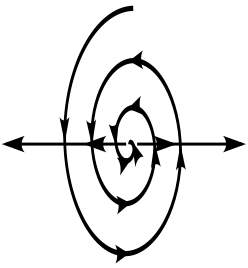}
\label{fig:As}
}
\quad
\subfigure[Type Bs]{
\includegraphics[width=1.47cm]{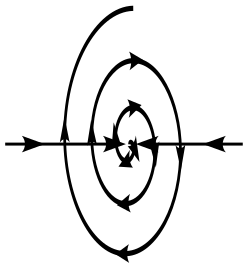}
\label{fig:Bs}
}
\caption{Classification of magnetic nulls -- Type A: real eigenvalues with signs (-,-,+); Type B: real eigenvalues with signs (+,+,-); Type As: two complex and one real eigenvalues with signs of real parts (-,-,+); Type Bs: two complex and one real eigenvalues with signs of real parts (+,+,-). Figure from~\cite{Franssila2023}.}\label{fig:NPclf} 
\vspace{-\baselineskip}
\end{figure}

Due to the above constraints, each null can be classified in terms of its \emph{polarity}; if the two same-sign eigenvalues are negative, the null is classified as a \emph{negative null}, otherwise it is a \emph{positive null} \cite{parnell2010}. Furthermore, magnetic nulls with complex eigenvalues exhibit a \emph{spiraling} topology \cite{longcope2005}. Figure~\ref{fig:NPclf} illustrates the resulting classification into four types, where types A and B represent (non-spiraling) topologies of negative and positive magnetic nulls, while types As and Bs encode spiraling topologies of negative and positive magnetic nulls, respectively.


\renewcommand\thesubfigure{}

\subsection{Separatrices and separators}
\label{sec: sec32}

\begin{figure*}[t]
    \centering
    \begin{minipage}{0.49\linewidth}
        \centering
        \includegraphics[width= 0.9\linewidth]{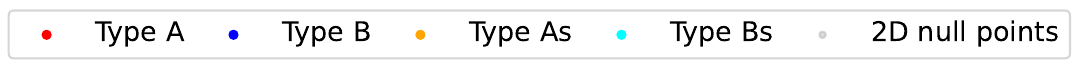}
        \includegraphics[width=0.95\linewidth]{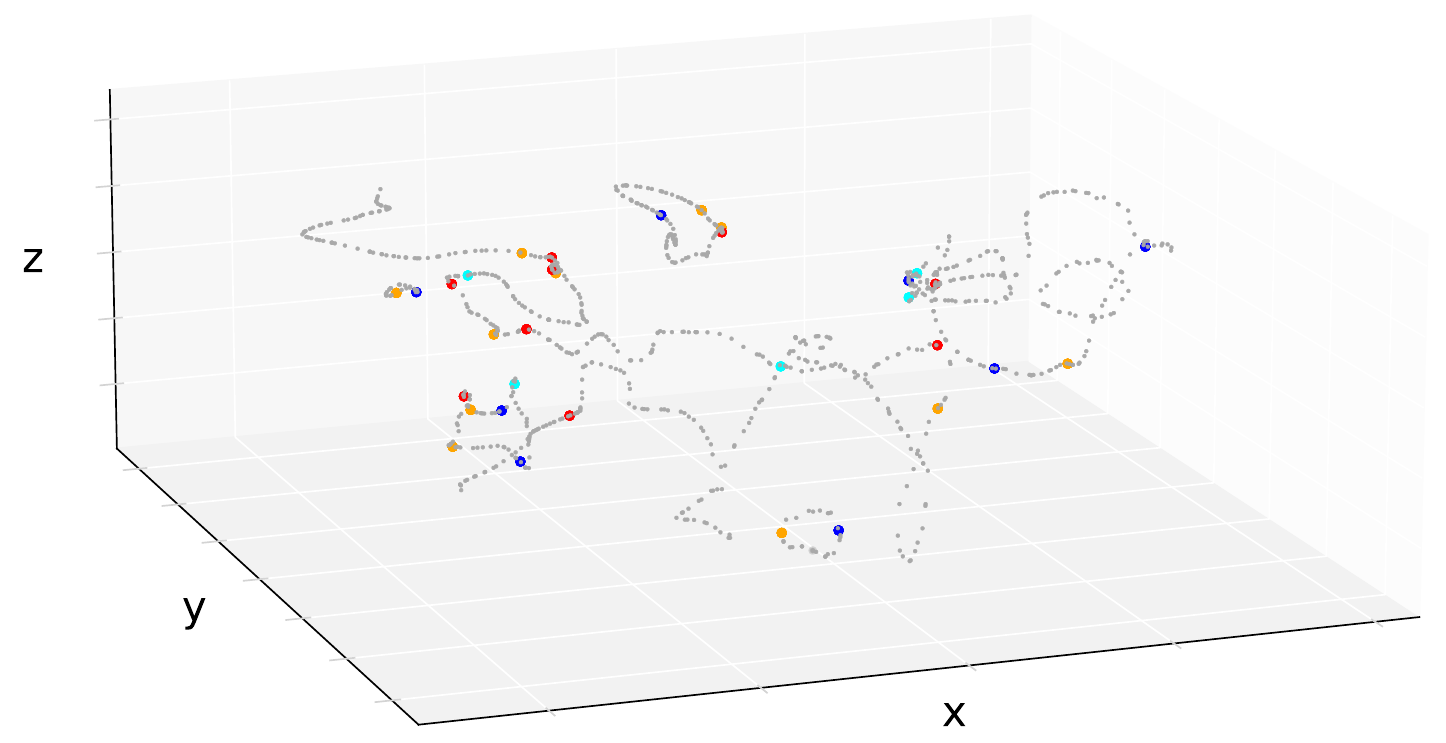}
    \end{minipage}
    \begin{minipage}{0.49\linewidth}
        \centering
        \vspace{0.2in}
        \includegraphics[width=0.95\linewidth]{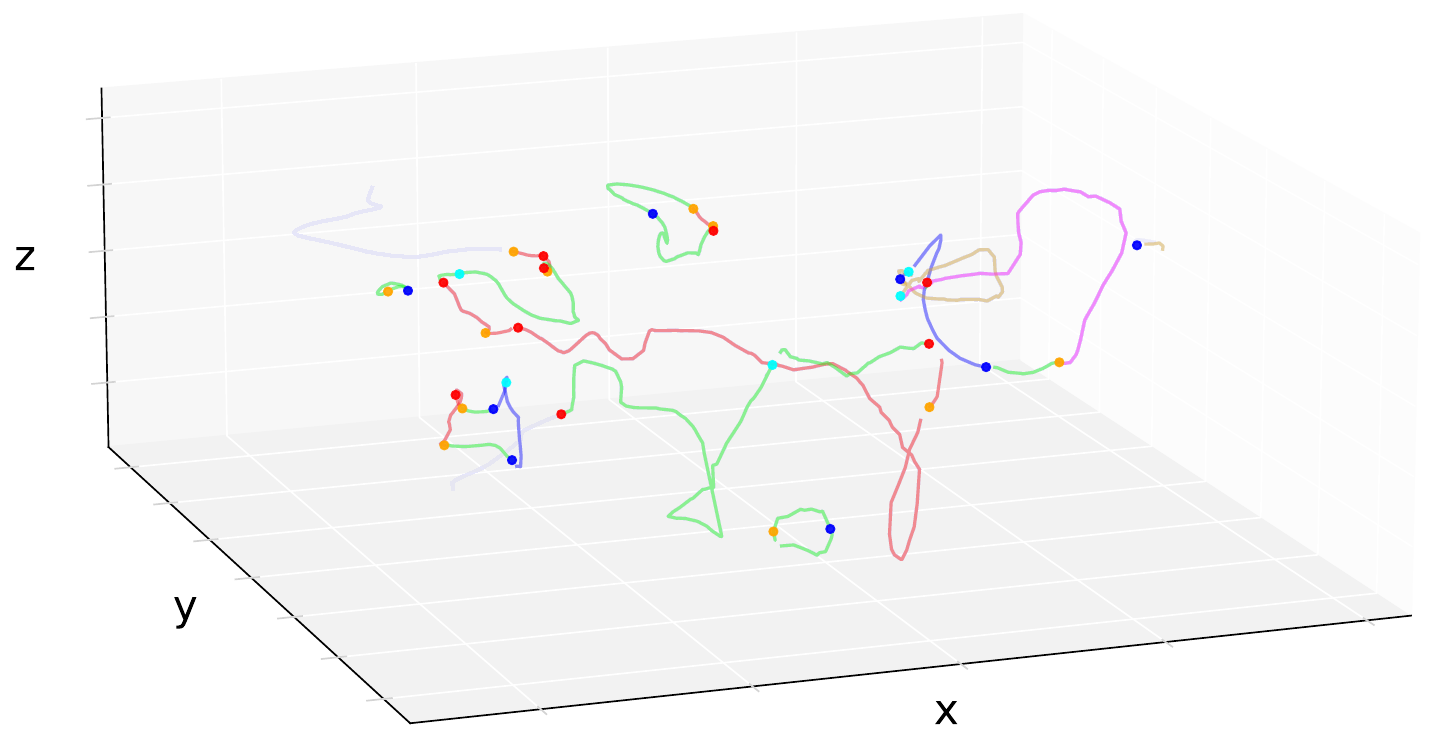}
    \end{minipage}
    \caption{\emph{Left:} Detected proper and 2D null points. \emph{Right:} The resulting spatial graph at the same time-step of the simulation. The spatial edges represent different types of separators in blue (As $\leftrightarrow$ Bs), red (A $\leftrightarrow$ B), green (A $\leftrightarrow$ Bs, B $\leftrightarrow$ As), pink (A $\leftrightarrow$ As, B $\leftrightarrow$ Bs) and yellow (A $\leftrightarrow$ A, B $\leftrightarrow$ B, As $\leftrightarrow$ As, Bs $\leftrightarrow$ Bs).}
    \vskip -0.15in 
    \label{fig:2}
\end{figure*}

The eigenvectors of the Jacobian can be used to define the so-called \emph{separatrices} that are associated with the magnetic nulls~\cite{longcope2005,Gun20b}. Each non-degenerate null has one 2D separatrix or \emph{fan surface} and two one-dimensional virtual separatrices or \emph{spines} \cite{longcope2005}. The fan surface is defined by the infinitely many magnetic field lines within the plane spanned by the two eigenvectors corresponding to the same-sign eigenvalues. The two spine field lines end in the magnetic null point, entering
along the directions parallel and antiparallel to the third eigenvector, normal to the fan plane \cite{longcope2005}.

In physical simulations, magnetic nulls connect via \emph{separator curves} (or reconnection lines) formed by the intersection of the fan surfaces of two connected nulls~\cite{palmroth_magnetopause_2006}. However, the process of integrating the separatrices to find their intersection can be computationally very expensive, which is why separators are usually approximated~\cite{haynes2010}.


In order to approximate the separators, we choose to follow the \emph{2D null lines}: curves along which two of the magnetic field components are zero, while the third can vary. \citet{zeiler2002} found that with a small guide-field, 3D reconnection is well-approximated by a 2D system. Specifically in the magnetotail environment, we ignore the East-West component of the magnetic field, $B_y$, since it tends to exhibit the least variation as the guide-field component, and we require that the other two components are zero, $B_x=B_z=0$. We call such points \emph{2D nulls} and use the term \emph{proper null} to refer to points where $\|\vec B\|_2=0$, which are clearly a subset of the 2D nulls. We provide a modification of the existing VTK vector field topology filter to detect such 2D nulls. 



\section{Graph Representations}

Originally introduced by \citet{longcope2002}, \emph{null graphs} are graph representations that characterize the topology of a magnetic field by encoding the connectivity between proper nulls (vertices) via separators (edges). 

We extend their definition to construct a graph representation that can be useful for different machine learning tasks and downstream applications, such as spatiotemporal GNNs \cite{reinhart2018}. We propose two computationally efficient heuristics to trace the connectivity between proper nulls both spatially and temporally.

\subsection{Spatial Null Graphs}
\label{sec:331}
After modifying the VTK vector field topology filter to detect the 2D magnetic nulls where $B_x = B_z = 0$ (sec.~\ref{sec: sec32}), we construct the 2D null lines by connecting the 2D nulls to each other based on spatial proximity. 



In practice, the 2D null lines can be traced by initializing at most two paths from each proper null based on a cut-off value on the maximum Euclidean distance from the proper null. 
Each of the paths is then iteratively expanded by finding -- within the same cut-off distance -- the nearest 2D null that is not already included in any of the already traced paths. Paths terminating without reaching a proper null are considered a dead end and are discarded, while paths ending at a proper null become edges in the null graph. The type of each proper null (A / B / As / Bs) is encoded as a node feature in the graph.

\subsection{Spatiotemporal Null Graphs}
\label{sec:332}

Consider a bipartite graph $\mathcal{G} = (\mathcal{V}_i, \mathcal{V}_{i+1}, \mathcal{E})$, where the vertex sets $\mathcal{V}_i$ and $\mathcal{V}_{i+1}$ are defined by proper nulls detected at times $t_i$ and $t_{i+1}$, respectively. The set of edges $\mathcal{E}$ represents a (partial) matching between the magnetic nulls with the interpretation that vertices $v\in\mathcal{V}_i$ and $v'\in\mathcal{V}_{i+1}$ are connected by an edge $e=(v,v')\in\mathcal{E}$ if they correspond to the same proper null. The problem can be cast as an \emph{unbalanced assignment problem} defined by the following maximization~problem
$$
\max_{\mathcal{E}\in\mathfrak{M}}
     \sum_{(v,v') \in \mathcal{E}} 1/w(v,v'),
$$
where the set of allowed matchings $\mathfrak{M}$ is defined by requiring that each vertex appears in at most one edge, and the weight $w(v,v') = \| \vec x(v) - \vec x(v')\|_2$ is defined by the Euclidean distance between the respective coordinates of the proper nulls $v$ and $v'$. Additional constraints including $(i)$ a maximum distance constraint $w_{\max}$, or $(ii)$ matching only the same type nulls, 
can be incorporated by letting $w(v,v')=-1$, for any edge that does not satisfy them. We apply both constraints to get an initial matching, and, for some of the unmatched magnetic nulls, we need to run a subsequent matching without constraint $(ii)$ to account for type switches.


All vertices in $\mathcal{V}_i$ that remain unmatched are considered to have disappeared after step $t_i$. Likewise, all vertices in $\mathcal{V}_{i+1}$ that remain unmatched are considered to have appeared before step $t_{i+1}$. According to \citet{murphy2015}, proper nulls can either appear by entering the simulation domain across a boundary, or as a result of a bifurcation, in which case they appear in pairs of opposite polarity. These two cases can be distinguished based on the coordinates and types of the unmatched vertices in $\mathcal{V}_{i+1}$.

\section{Results and discussion}

The supercomputer-generated simulations from Vlasiator provide large-scale, high-fidelity data -- some of which is openly available \cite{vlasiator_opendata}. To give an idea of the scale of available data, the example of the published Vlasiator dataset provides 170 time-steps $\times$ 1,723,328 grid points at each time-step, i.e., the time series consists of a total of $\sim$~293~million~grid~points. Working with such scales requires efficient and scalable methods, and the spatiotemporal graph representation of the data can allow for less resource-intensive machine learning approaches.

The magnetotail simulations used to produce the results presented here, consist of a grid with resolution $50 \times 127 \times 108$ in the $x$ (tailward--Earthward), $y$ (East--West), and $z$ (North--South) directions of the magnetic field, respectively. A step of one unit in any direction of the grid corresponds to 1000~km, and the time series used to generate the spatiotemporal null graphs has 1~s cadence \cite{Franssila2023}.

First, the modified VTK topology filter is used to detect 2D and proper nulls, and to classify the latter in types (sec.~\ref{sec:sec31}). The 2D nulls are detected using $B_y$ as the guide-field component (sec.~\ref{sec: sec32}). 
The results obtained from the first stage of the process are illustrated on the left side of Figure~\ref{fig:2}. 

The result of the spatial tracing method (sec.~\ref{sec:331}), is presented on the right of Figure \ref{fig:2}. The 3D null points are colored according to their type. Different types of spatial connectivity are also color-encoded, using different colors of spatial edges depending on the type of the connection. 

\begin{figure}[t]
    \centering
    \includegraphics[width=0.99\columnwidth]{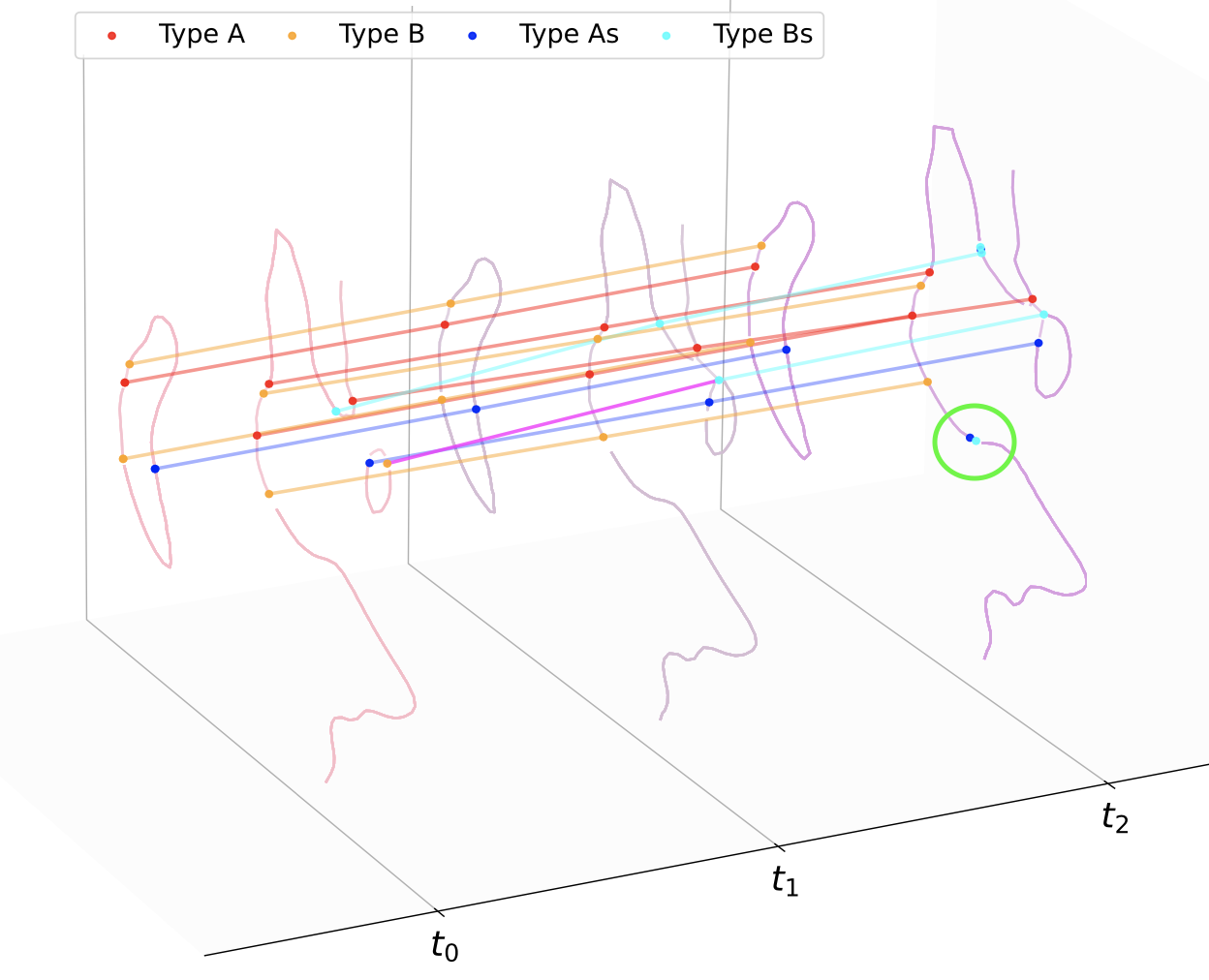}
    \caption{Spatiotemporal null graph. At each time-step a 2D projection (X-Y plane) of the corresponding spatial graph is illustrated. The proper nulls are colored based on their type, and the color of each temporal edge represents the type of the traced proper null, with the exception of the pink edge which denotes a type switch. The green circle at $t_2$ marks a pair of proper nulls of opposite polarity that appear together after a bifurcation.}
    \vspace{-\baselineskip}
    \label{fig4}
\end{figure}

Figure \ref{fig4} shows an example of a spatiotemporal graph, where for each time-step $t_i$ for $i \in \{0,1,2\} $, a 2D projection (X-Y plane) of the spatial graph is presented. The temporal edges trace the temporal evolution of each proper null across all time-steps. The colors of the temporal edges represent the type of the proper null traced over time, with the exception of the pink temporal edges which denote a type switch scenario (e.g., $t_0 \rightarrow t_1$: B $\rightarrow$ Bs). Finally, the green circle at $t_2$ is used to mark a pair of proper nulls of opposite polarity that appear together before $t_2$ due to a bifurcation.





We have presented a scalable data analysis pipeline for the detection and spatiotemporal tracing of proper magnetic nulls. These methods allow us to characterize the topology a 3D magnetic field using graph representations. The resulting spatiotemporal null graphs can be useful in various downstream learning tasks, especially in GNN applications \cite{reinhart2018, zhou2020, wu2020}.

In the process of formulating 3D magnetic reconnection detection as a machine learning task, two potential limitations arise. If we formulate the problem as a supervised learning task, there is a severe difficulty in reliably labeling a sufficient amount of training data, as 3D magnetic reconnection remains difficult to detect and characterize. Similarly, if we were to formulate the problem as an unsupervised learning task, questions arise regarding the interpretability of results and the model performance evaluation.

Currently, we are working on a GNN approach that aims to circumvent these issues by formulating the learning task as a plasmoid\footnote{Outflows of plasma driven by the magnetic tension force of newly reconnected field lines \cite{liu2013}.} formation forecast, as their generation is linked to reconnecting plasmas \cite{samtaney2009}. The location of a plasmoid can be characterized using the magnetic skeleton \cite{birn1997}, which allows us to use spatiotemporal null graphs to learn when and where a plasmoid is formed. Next, in order to detect the magnetic reconnection, we can examine the reconnection rate and energy conversion rate at the possible reconnection sites located in close proximity to the newly-formed plasmoid. This work can then be extended to facilitate the study of magnetized plasmas in different settings, which is linked to a variety of open questions that can be interesting to both the astrophysics and machine learning research communities. 

\section*{Acknowledgements}
Funding in direct support of this work: Research Council of Finland grants \#345635 (DAISY) and \#339327 (Carrington). The authors thank the Finnish Computing Competence Infrastructure (FCCI) for supporting this work with computational and data storage resources.


\bibliography{example_paper}
\bibliographystyle{icml2023}

\newpage
\appendix
\onecolumn


\end{document}